\documentclass[reprint,aps,prl,showpacs,floatfix,superscriptaddress,citeautoscript,longbibliography]{revtex4-1}

\pdfoutput=1

\usepackage{amsmath, amsfonts, amssymb, amsbsy}
\usepackage{graphicx}
\usepackage{xspace}

\usepackage[margin=19mm]{geometry}

\usepackage[caption=false]{subfig}

\makeatletter
\def\tagform@#1{\maketag@@@{\ignorespaces#1\unskip\@@italiccorr}}
\makeatother

\usepackage{hyperref}
\hypersetup{
        pdftitle={Direct observation and temperature control of the surface Dirac gap in the topological crystalline insulator (Pb,Sn)Se},
        pdfdisplaydoctitle=true,
        pdfsubject = {Article},
        pdfauthor= {BMW},
        pdfkeywords= {},
        colorlinks=true,
        linkcolor=blue,
        citecolor=blue,
        urlcolor=blue,
        unicode=false
}

\newcommand{\PbSnXSe}{Pb\texorpdfstring{$_{1-x}$}{\ifpdfstringunicode{\unichar{"2081}\unichar{"208B}\unichar{"2093}}
{1-x}}Sn\texorpdfstring{$_{x}$}{\ifpdfstringunicode{\unichar{"2093}}{x}}Se\xspace}

\begin{document}
\title{Direct observation and temperature control of the surface Dirac gap in the topological crystalline insulator (Pb,Sn)Se}

\author{B.~M. Wojek}
\homepage{http://bastian.wojek.de/}
\affiliation{KTH Royal Institute of Technology, ICT MNF Materials Physics, Electrum 229, 164 40 Kista, Sweden}
\author{M.~H. Berntsen}
\affiliation{KTH Royal Institute of Technology, ICT MNF Materials Physics, Electrum 229, 164 40 Kista, Sweden}
\author{V.~Jonsson}
\affiliation{KTH Royal Institute of Technology, ICT MNF Materials Physics, Electrum 229, 164 40 Kista, Sweden}
\affiliation{Center for Quantum Materials, Nordic Institute for Theoretical Physics (NORDITA), Roslagstullsbacken 23, 106 91 Stockholm, Sweden}
\author{A.~Szczerbakow}
\affiliation{Institute of Physics, Polish Academy of Sciences, Aleja Lotnik\'{o}w 32/46, 02-668 Warsaw, Poland}
\author{P.~Dziawa}
\affiliation{Institute of Physics, Polish Academy of Sciences, Aleja Lotnik\'{o}w 32/46, 02-668 Warsaw, Poland}
\author{B.~J. Kowalski}
\affiliation{Institute of Physics, Polish Academy of Sciences, Aleja Lotnik\'{o}w 32/46, 02-668 Warsaw, Poland}
\author{T.~Story}
\affiliation{Institute of Physics, Polish Academy of Sciences, Aleja Lotnik\'{o}w 32/46, 02-668 Warsaw, Poland}
\author{O.~Tjernberg}
\email{oscar@kth.se}
\affiliation{KTH Royal Institute of Technology, ICT MNF Materials Physics, Electrum 229, 164 40 Kista, Sweden}
\affiliation{Center for Quantum Materials, Nordic Institute for Theoretical Physics (NORDITA), Roslagstullsbacken 23, 106 91 Stockholm, Sweden}

\date{\today}

\begin{abstract}
Since the advent of topological insulators hosting symmetry-protected Dirac surface states, efforts have been made to gap these states in a controllable way. A new route to accomplish this was opened up by the discovery of topological crystalline insulators (TCIs) where the topological states are protected by real space crystal symmetries and thus prone to gap formation by structural changes of the lattice. Here, we show for the first time a temperature-driven gap opening in Dirac surface states within the TCI phase in (Pb,Sn)Se. By using angle-resolved photoelectron spectroscopy, the gap formation and mass acquisition is studied as a function of composition and temperature. The resulting observations lead to the addition of a temperature- and composition-dependent boundary between massless and massive Dirac states in the topological phase diagram for (Pb,Sn)Se (001). Overall, our results experimentally establish the possibility to tune between a massless and massive topological state on the surface of a topological system.
\end{abstract}


\maketitle

\section{Introduction}
\label{sec:intro}

\begin{figure}
	\centering
	\includegraphics[width=.8\columnwidth]{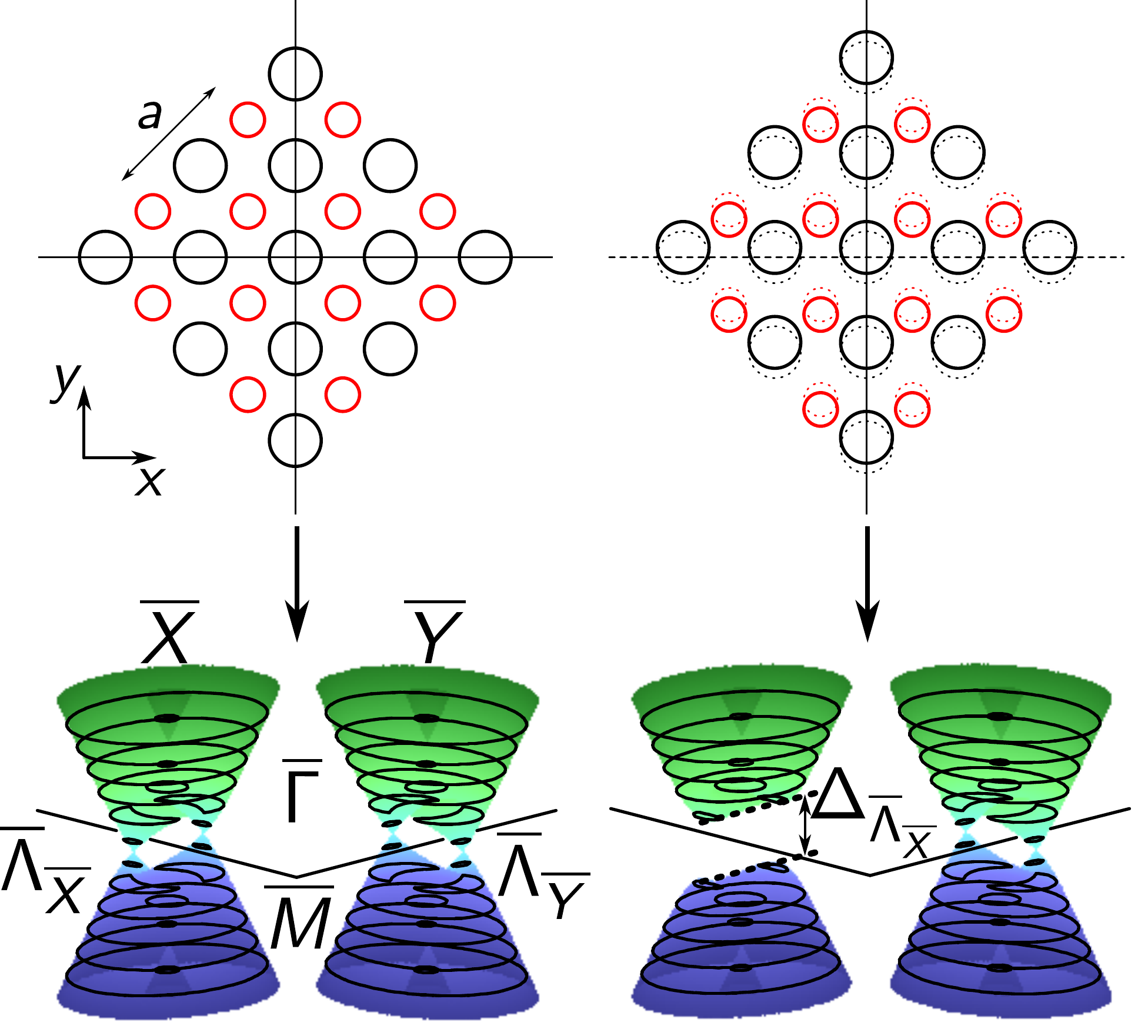}
	\caption{Sketch of the (001) surface of (Pb,Sn)Se and the corresponding surface band structures in the TCI state. The particle-hole-symmetric band structure consists of the well-known hybridized Dirac cones, featuring a Lifshitz transition indicated by the constant-energy contours. The illustrations to the left show the situation of an undistorted surface hosting only gapless surface states. The picture to the right shows a surface in which the cations and anions have shifted with respect to each other along one of the mirror lines as suggested by STM measurements~\cite{Zeljkovic-NatMater-2015}. The breaking of the second mirror symmetry induces a gap in one pair of the surface states. In this work, $x$ and $y$ (and $k_x$ and $k_y$, correspondingly) are defined \emph{in the surface} as shown. These coordinates are rotated by $45^{\circ}$ about $z$ with respect to the conventionally used bulk coordinates.}
	\label{fig:sketch}
\end{figure}

The study of topological properties of materials and corresponding phase transitions has received tremendous interest in
the condensed-matter community in the recent years~\cite{Hasan-RevModPhys-2010, Ando-JPSJ-2013}. A particularly interesting
class of materials are three-dimensional topological crystalline insulators (TCIs)~\cite{Fu-PhysRevLett-2011, AndoFu-AnnRevCMP-2015}
where degeneracies in the surface electronic band structures are protected by point-group symmetries of the crystals. The
first material systems to realize this state were found within the class of IV-VI narrow-gap semiconductors:
SnTe~\cite{Hsieh-NatCommun-2012, Tanaka-NatPhys-2012}, (Pb,Sn)Te~\cite{Xu-arXiv-2012, Tanaka-PhysRevB-2013}, and
(Pb,Sn)Se~\cite{Dziawa-NatMater-2012, Wojek-PhysRevB-2013}. These materials, crystallizing in the rock-salt structure, host a set of
four Dirac points on specific surfaces and a great advantage in particular of the solid solutions is the tunability of the bulk
band gap---and hence their topological properties---by several independent parameters~\cite{NimtzAndSchlicht}, such as
composition~\cite{Tanaka-PhysRevB-2013, Wojek-PhysRevB-2014}, temperature~\cite{Dziawa-NatMater-2012, Wojek-PhysRevB-2013, Wojek-PhysRevB-2014},
pressure~\cite{Barone-PhysRevB-2013}, or strain~\cite{Barone-PhysStatSolidi-2013}. Moreover, since the surface Dirac points are protected
by crystalline symmetries, it has been predicted that selectively breaking these symmetries can lift the surface-state degeneracies
and the carriers acquire masses~\cite{Hsieh-NatCommun-2012, SerbynFu-PhysRevB-2014}. Indeed, signatures of such massive surface Dirac fermions
have been observed recently in Landau-level-spectroscopy experiments on (001) surfaces of (Pb,Sn)Se crystals~\cite{Okada-Science-2013} and
a surface distortion has been identified to be at the heart of the phenomenon~\cite{Zeljkovic-NatMater-2015} (cf. Fig.~\ref{fig:sketch}). However, since these scanning-tunneling-microscopy/spectroscopy (STM/STS) studies only provide low-temperature data, the evolution of the surface-state mass/gap with temperature remains elusive. It is particularly interesting to ask whether the four (ungapped) Dirac points ever exist simultaneously on the (001) surface of (Pb,Sn)Se crystals in the TCI state, whether the partially gapped surface states can be observed directly by $k$-resolved methods, and whether one can switch between the massive and massless states by varying extrinsic parameters.

In this study, we address these open questions using high-resolution angle-resolved photoelectron spectroscopy (ARPES) measurements on (Pb,Sn)Se mixed crystals. Consistent with the previous STM/STS studies we directly observe robust gapped Dirac states on the (001) surface at low temperature. Most importantly, however, the evolution of the gap/mass is tracked by composition- and temperature-dependent experiments and a transition to entirely massless surface Dirac states is witnessed. We suggest that the surface distortion detected by STM vanishes at sufficiently high temperatures, thus leading to the possibility of tuning the surface states employing structural changes. Subsequently, this is summarized in a revised topological phase diagram for (Pb,Sn)Se (001).

\section{Experimental details and results}

The $n$-type (Pb,Sn)Se single crystals with typical doping levels of a few $10^{18}$~cm$^{-3}$ studied in this work have been grown using the self-selecting vapor-growth method and characterized by X-ray diffraction and energy-dispersive X-ray spectroscopy~\cite{Szczerbakow-JCrystalGrowth-1994, *Szczerbakow-ProgCrystalGrowth-2005}. High-resolution ARPES studies were carried out on samples cleaved in a (001) plane in ultra-high vacuum at room temperature. Temperature-dependent measurements of solid solutions with SnSe/PbSe ratios between $0$ and $0.6$ were carried out at the BALTAZAR laser-ARPES facility using linearly polarized light with an excitation energy $h\nu=10.5$~eV and a THEMIS-1000 time-of-flight analyzer collecting the photoelectrons~\cite{Berntsen-RevSciInstrum-2011}. The overall crystal-momentum and energy resolution is better than $0.008$~\AA{}$^{-1}$ and about $5$~meV, respectively.

\begin{figure}
	\centering
	\subfloat{\includegraphics[width=\columnwidth]{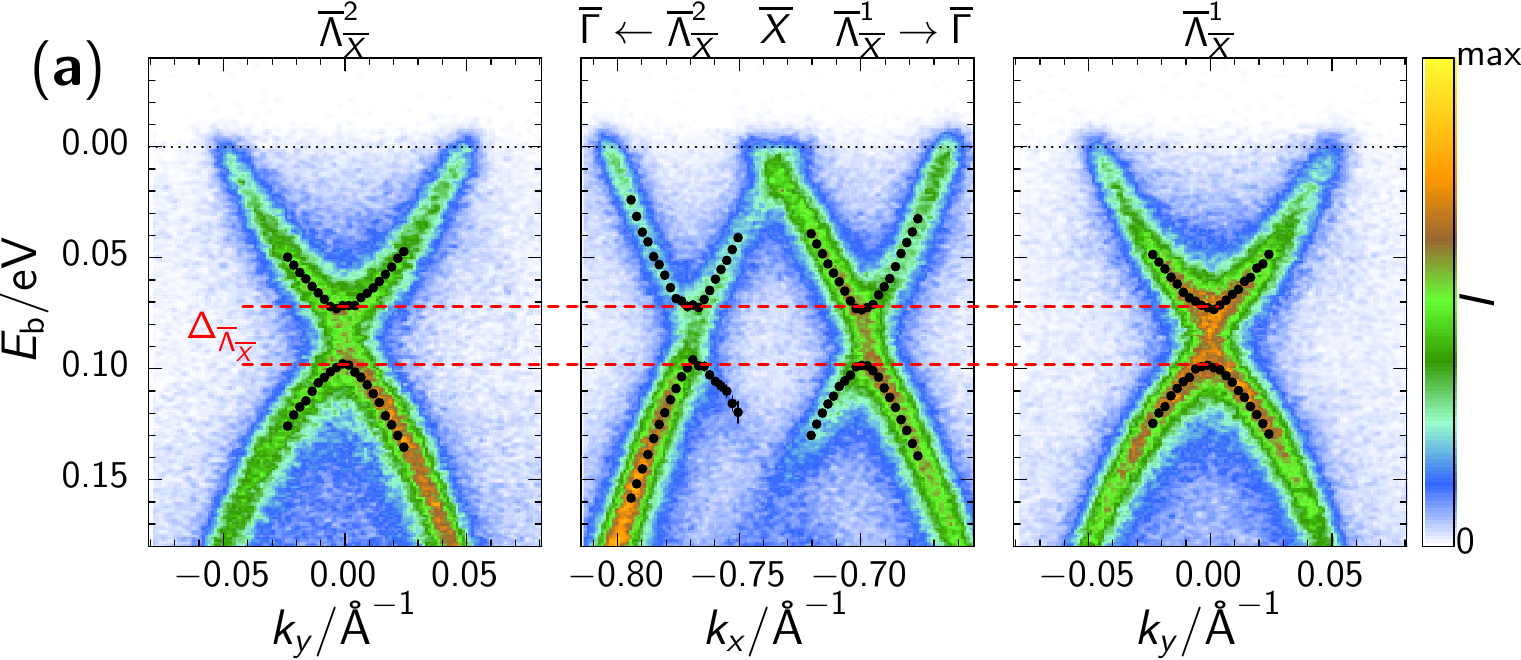}\label{fig:gap:a}}\\
   	\subfloat{\includegraphics[width=0.495\columnwidth]{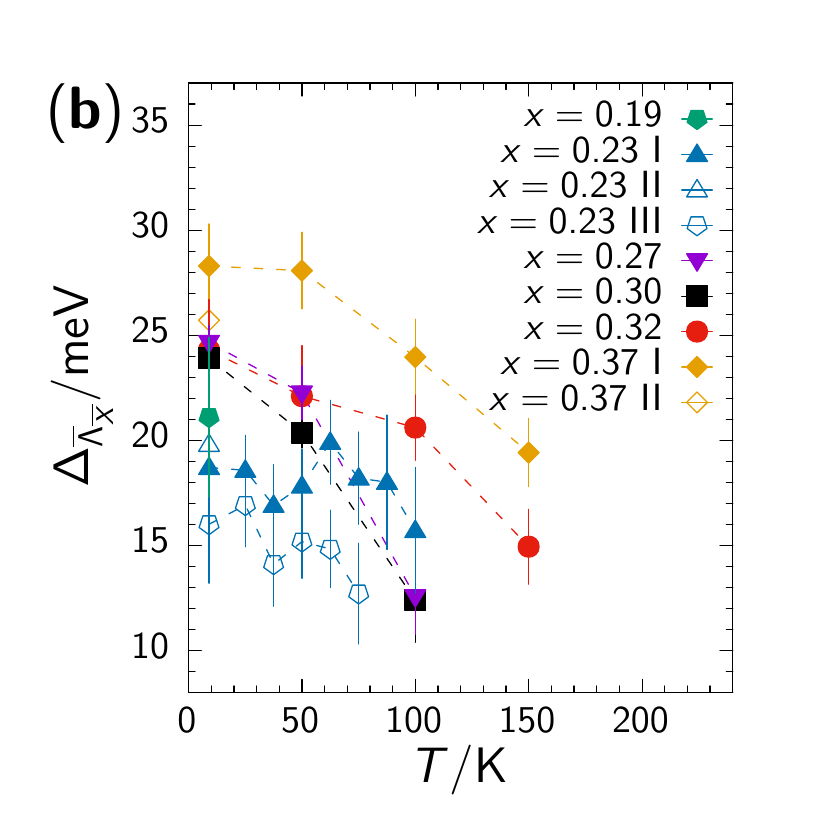}\label{fig:gap:b}}\hfill
   	\subfloat{\includegraphics[width=0.495\columnwidth]{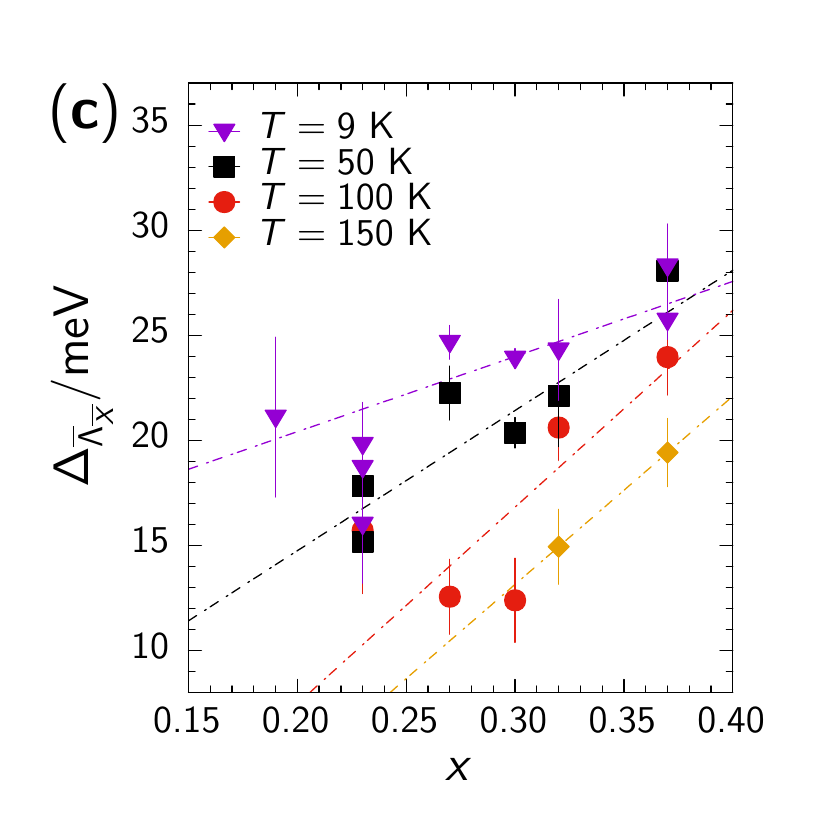}\label{fig:gap:c}}
	\caption{Surface-state Dirac gap in (Pb,Sn)Se (001). (a) ARPES spectra showing the surface-state dispersion of Pb$_{0.63}$Sn$_{0.37}$Se in the vicinity of $\overline{X}$ at $T=9$~K. The reciprocal-space cuts are taken parallel to the borders of the surface Brillouin zone through the ``Dirac points'' at $\Lambda_{\overline{X}}$. The black circles represent the obtained peak positions from a two-Voigtian fit to the energy distribution curves. A gap $\Delta_{\overline{\Lambda}_{\overline{X}}}$ of about $25$~meV is observed consistently across the spectra. (b) Surface-state gap  $\Delta_{\overline{\Lambda}_{\overline{X}}}$ as a function of temperature for various cleaved (001) surfaces of \PbSnXSe. (c) $\Delta_{\overline{\Lambda}_{\overline{X}}}$ as a function of SnSe content $x$ for selected temperatures. The dash-dotted lines indicate the rough trend in the observed gap size.}
	\label{fig:gap}
\end{figure}

While studying the detailed nature of the bulk band inversion in (Pb,Sn)Se~\cite{Wojek-PhysRevB-2014}, low-temperature data of certain samples with comparably high SnSe content were found to contain a peculiar distribution of spectral weight around the Dirac points. An example of such spectra (not included in Ref.~\onlinecite{Wojek-PhysRevB-2014}) is depicted in Fig.~\ref{fig:gap:a}. The characteristic band structure of the hybridized ``parent surface Dirac cones'' schematically shown in Fig.~\ref{fig:sketch} is clearly observed close to the $\overline{X}$ point of the (001) surface Brillouin zone. However, also a distinct gap opening at the Dirac nodes on the $\overline{\Gamma}$-$\overline{X}$ line is revealed. Following this initial finding and reports of similar observations using STM/STS techniques~\cite{Okada-Science-2013}, the gap formation has been studied systematically across the available temperature and composition ranges.

\begin{figure*}
	\centering
	\begin{minipage}[c]{0.71\textwidth}
		\vspace{0pt}\subfloat{\includegraphics[width=\textwidth]{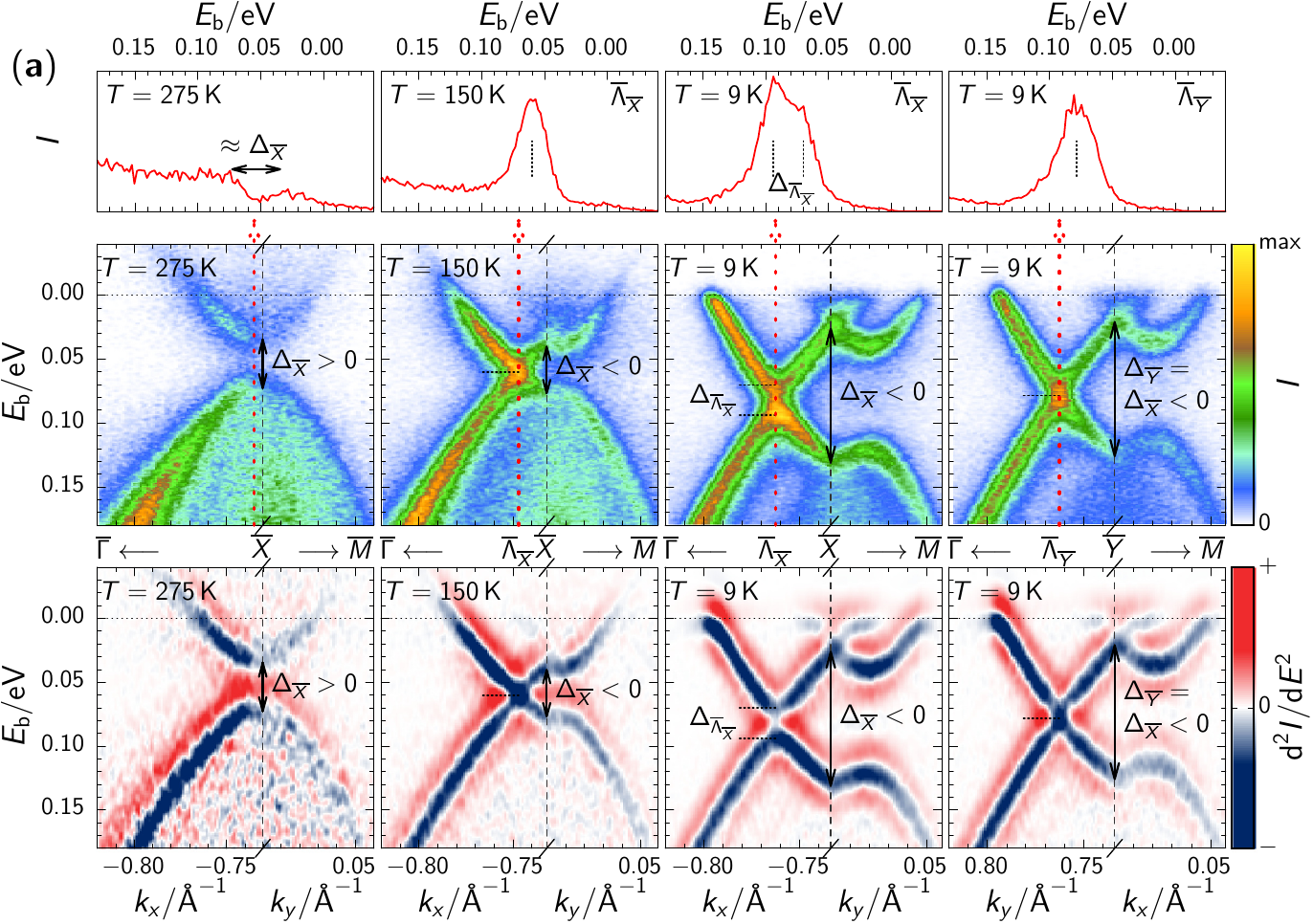}\label{fig:phasediagram:a}}
	\end{minipage}
	\hfill
	\begin{minipage}[c]{0.28\textwidth}
		\vspace{0pt}\subfloat{\includegraphics[width=\textwidth]{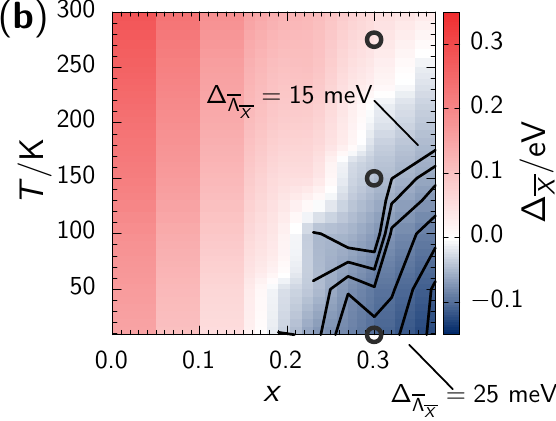}\label{fig:phasediagram:b}}\\
		\subfloat{\includegraphics[width=.9\textwidth]{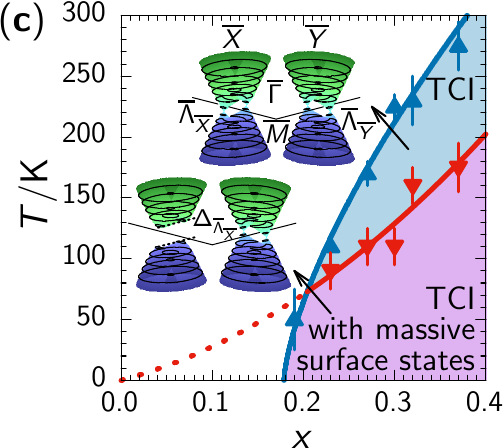}\label{fig:phasediagram:c}}
	\end{minipage}
	\caption{Topological transitions in (Pb,Sn)Se (001). (a) ARPES data of Pb$_{0.70}$Sn$_{0.30}$Se (001) at different temperatures in the vicinity of $\overline{X}$/$\overline{Y}$: the graphs in the center depict the absolute intensity along the high-symmetry lines of the surface Brillouin zone, the ones on the bottom show the second derivative with respect to the energy of the data, and the plots in the top row present EDCs along the red dotted lines in the spectra. (b) Summary of the evolutions of the bulk gap [$\Delta_{\overline{X}}$, color map, cf. Ref.~\onlinecite{Wojek-PhysRevB-2014}] and the largest observed surface Dirac gap [$\Delta_{\overline{\Lambda}_{\overline{X}}}$, contour lines, spacing $2.5$~meV, cf. Figs.~\ref{fig:gap:b} and~\ref{fig:gap:c}] in \PbSnXSe (001). The open gray circles indicate the ($x$, $T$) positions of the data shown in (a). (c) Topological phase diagram for (Pb,Sn)Se (001) based on the observations of the spectral gaps shown in (b). All lines are guides to the eye. The dotted line marks the proposed transition line for the occurrence of the surface distortion in the topologically trivial state.}
	\label{fig:phasediagram}
\end{figure*}

The decision whether the surface states in the TCI phase are massive/gapped is based on an analysis of the energy-distribution curves (EDCs) in the close vicinity of the Dirac nodes ($\overline{\Lambda}$). In a first step, the EDCs are modeled by sums of two Voigtian lines on a small strictly monotonic linear to cubic background. The resulting line widths are compared to a fit of a single Voigtian (on the same background) to the EDC at $\overline{\Lambda}$. In case the single line (full width at half maximum) at $\overline{\Lambda}$ is found to be broader by more than $5$~meV than the broadest two-line constituent, we call the states gapped and the two-line peak separation determines the gap size $\Delta_{\overline{\Lambda}}$~\cite{5meV}. Using this criterion, we find most low-temperature spectra are gapped. Yet, at higher temperatures the Dirac nodes are---within the resolution of the experiment---overall intact. Figures~\ref{fig:gap:b} and~\ref{fig:gap:c} summarize the ascertained gap sizes for the samples investigated during this study. While one has to note that the values of $\Delta_{\overline{\Lambda}}$ vary by about $3$~meV to $4$~meV for nominally similar sample compositions, nevertheless, a general trend to larger values for higher SnSe contents and lower temperatures is apparent.

Having directly established the low-temperature mass acquisition of the Dirac fermions at $\overline{\Lambda}_{\overline{X}}$ it remains to be investigated whether the Dirac points at $\overline{\Lambda}_{\overline{Y}}$ do exist also on the distorted surface as suggested by the STM/STS studies~\cite{Okada-Science-2013, Zeljkovic-NatMater-2015}. Unlike STM/STS which is a local-probe technique, ARPES averages over a lateral sample region with a diameter of several tens of micrometers. Hence, \emph{a priori} the observation of both massive surface states at $\overline{\Lambda}_{\overline{X}}$ and massless states at $\overline{\Lambda}_{\overline{Y}}$ requires a long-range textured surface. If rather a multi-domain structure is formed with distortions along $x$ and $y$, respectively for different domains, a superposition of gapped and gapless states is expected to be seen. A further complication arises from the low photon energy and the experimental geometry of the ARPES set-up which necessitate the samples under study to be rotated by $\pm 90^{\circ}$ to enable the collection of electrons originating from $\overline{Y}$. Since most likely laterally slightly different parts of the surface are probed before and after such a manipulation, the long-range-texture precondition is further intensified. Our experiments show that this requirement is not always met (and hence, overall massive Dirac states are observed). Nevertheless, indeed we do find Dirac nodes, and thus massless states, at $\overline{\Lambda}_{\overline{Y}}$ coexisting with gapped states at $\overline{\Lambda}_{\overline{X}}$ at low temperatures. An example of the latter is shown in Fig.~\ref{fig:phasediagram:a}. It illustrates the evolution of the (001) surface states of Pb$_{0.70}$Sn$_{0.30}$Se across the different phases. Close to room temperature the sample is in a topologically trivial state characterized by the positive bulk gap at $\overline{X}$ ($L$)~\cite{Wojek-PhysRevB-2014}. Upon cooling to $T=150$~K, the bulk bands invert and the sample enters the TCI phase with massless surface states. Lowering the temperature further leads to the gap formation at the positions of the former Dirac points $\overline{\Lambda}_{\overline{X}}$, while the degeneracies at $\overline{\Lambda}_{\overline{Y}}$ prevail, thus confirming the overall situation depicted in Fig.~\ref{fig:sketch}.

\section{Discussion}

In order to understand the full implications of this study, it is necessary and helpful to compare our results with the observations made in STM/STS experiments before. Zeljkovic \emph{et al.}~\cite{Zeljkovic-NatMater-2015} have demonstrated a low-temperature ($T=4$~K) surface Dirac gap in (Pb,Sn)Se reaching a value of about $25$~meV at a SnSe content of about $38$~\% in the solid solution. With decreasing $x$ the gap is reported to decrease and it supposedly vanishes at a critical composition of about $17$~\%. At the same time, the observed lattice distortion at the surface appears to have roughly the same magnitude for all studied samples. To reconcile the diminishing surface gap, it has been argued that the penetration depth of the surface state into the bulk increases when the size of the bulk gap decreases and thus the weight of the surface state in the outermost atomic layer probed by STM/STS decreases continuously. So, is our observation of the ($x$, $T$)-dependent gap merely the result of the varying surface-state penetration depth ($\propto \Delta_{\overline{X}}^{-1}$)? The answer to that question is found in Figs.~\ref{fig:gap:b} and~\ref{fig:gap:c}. At $T=9$~K, $\Delta_{\overline{\Lambda}_{\overline{X}}}$ indeed decreases with decreasing $x$, but in total only by a few millielectronvolts from $x=0.37$ to $x=0.19$. A similar behavior is found at $x=0.23$: Upon increasing the temperature from $9$~K to $100$~K (spanning essentially the entire band-inverted region) the observed gap is reduced only minutely. Hence, $\Delta_{\overline{\Lambda}_{\overline{X}}}$ observed here is---if at all---only to a small extent influenced by the changing penetration depth of the surface states. Thus, possibly due to the increased probing depth of low-energy ARPES as compared to STM/STS (nanometers instead of \aa{}ngstr\"oms), the gap can rather be regarded as a direct measure of the size of the lattice distortion at the surface~\cite{SerbynFu-PhysRevB-2014}. This conclusion is seemingly at variance with the claims of Ref.~\onlinecite{Zeljkovic-NatMater-2015}, yet, it should be noted that in the STM/STS work the reliably determined gap sizes span a region between about $15$~meV and $30$~meV, in full agreement with our data. The nonobservation of distinct features in the STS spectra due to the diminishing spectral weight in the top-most atomic layer does not imply an overall vanishing surface-state gap for samples close to the transition.

Altogether, our observations are summarized in Fig.~\ref{fig:phasediagram:b}. Based on the distinction of the massive and massless surface states in the bulk-band-inverted region we are able to determine a ``transition line'' below which the lattice distortion at least in the (001) surface occurs [Fig.~\ref{fig:phasediagram:c}]. Given the finite resolution of our experiment, it should be regarded as a lower limit and hence the indicated TCI phase with completely massless surface states might occupy a somewhat smaller, yet still finite, area in the ($x$, $T$) phase diagram. Since STM also provided evidence for a lattice distortion in the topologically trivial phase at low temperature~\cite{Zeljkovic-NatMater-2015}, although inaccessible by our spectroscopic method, we suspect the transition to occur at finite temperatures in this phase of the solid solutions as well.

The (Pb,Sn)Se mixed crystals with a SnSe content in the studied range are found to crystallize in the rock-salt structure without bulk structural phase transitions. Nevertheless, the occurrence of a surface distortion further corroborated by this study should not entirely catch us by surprise. SnSe itself features an orthorhombic crystal structure~\cite{Wiedemeier-ZKri-1978} and various closely related IV-VI solid solutions undergo ferroelectric phase transitions~\cite{Lebedev-Ferroelectrics-1994}. Also, in (Pb,Sn)Se the transverse-optical phonon mode at $\Gamma$ softens as the temperature is lowered and the SnSe content is increased~\cite{Vodopyanov-JETPLett-1978}. Occasionally, for Pb$_{0.59}$Sn$_{0.41}$Se even bulk structural changes were reported and found to be suppressed only with increasing carrier densities~\cite{Volkov-JETPLett-1978}, suggesting displacive transitions driven by electron-phonon coupling~\cite{Volkov-JETP-1978, Konsin-Ferroelectrics-1982}.

It is too early to conclude on the exact mechanism which leads to the observed surface structure. Specifically, given the small but still clearly detectable variations between different measured samples with comparable SnSe content, it appears that the local nature of this lattice distortion somewhat depends on the details of each surface. However, its overall tunability by independent parameters inferred from the ($x$, $T$) phase diagram [Fig.~\ref{fig:phasediagram:c}] lays the foundation for achieving a controlled mass generation in the topological surface Dirac states. It will be interesting to see in the future how other TCI surfaces behave~\cite{Safaei-PhysRevB-2013, Liu-PhysRevB-2013}, both of single crystals and of thin films~\cite{Yan-PhysRevLett-2014, Polley-PhysRevB-2014}, where controlled tuning of the crystal structure, e.g. by strain, promises to be a viable avenue to open gaps in topological surface states in addition to breaking symmetries by electric~\cite{Liu-NatMater-2014} or magnetic fields~\cite{SerbynFu-PhysRevB-2014}.

Eventually, it is worth noting that by specifically breaking both mirror symmetries as well as the rotational symmetries of the surface, all Dirac points are expected to be gapped in the material studied here~\cite{SerbynFu-PhysRevB-2014}. If a domain structure is formed, a topologically protected 1D conducting state percolating through the surface along the domain walls would be the result~\cite{Hsieh-NatCommun-2012}. Such states can potentially be engineered in very thin films grown on specific vicinal surfaces or by employing TCIs naturally featuring suitable lattice distortions.

\section{Summary}

In summary, the evolution of the surface-state gap/mass in (Pb,Sn)Se (001) has been determined systematically by means of high-resolution ARPES. Our results are consistent with the proposed underlying surface distortion at low temperatures. The observation of only massless Dirac surface states at higher temperatures indicates a structural transition of the surface. While its detailed nature is yet to be investigated by systematic structural and/or vibrational studies, we have shown experimentally the possibility to tune between massive and gapless topological states using this very transition.

\begin{acknowledgments}
We thank S.~S. Pershoguba and A.~V. Balatsky for stimulating discussions. This work was made possible through support from the Knut and Alice Wallenberg Foundation, the Swedish Research Council, and the Polish National Science Centre (NCN) grants No. 2011/03/B/ST3/02659 and 2012/07/B/ST3/03607.
\end{acknowledgments}


\begin{thebibliography}{32}
\bibitem [{5me()}]{5meV}%
\BibitemOpen
\href@noop {} {}\bibinfo {note} {The criterion is based on the scattering of
	the determined line width at high temperatures and the energy resolution of
	the experiment which are both on the order of $5$~meV.}\BibitemShut {Stop}%
\end{thebibliography}
\end{document}